\documentclass[12pt]{iopart}

\expandafter\let\csname equation*\endcsname\relax
\expandafter\let\csname endequation*\endcsname\relax
\usepackage{amsmath}
\usepackage{amssymb}
\usepackage{graphicx}
\usepackage[dvipsnames]{xcolor}
\usepackage{cite}

\usepackage{bm}

\newcommand{\m}[1]{\langle{#1}\rangle}

\newcommand{\dl}{\delta\ell}
\newcommand{\df}{\delta F}

\usepackage{siunitx}
\newcommand{\ppf}{\textit{Physarum polycephalum}}
\newcommand{\pp}{\textit{P.~polycephalum}}

\begin{document}

\title[Vascular adaptation model]{Vascular adaptation model from force balance: \\ \ppf~as a case study}

\author{Sophie Marbach}
 \address{CNRS, Sorbonne Universit\'{e}, Physicochimie des Electrolytes et Nanosyst\`{e}mes Interfaciaux, F-75005 Paris, France}
  \address{Courant Institute of Mathematical Sciences, New York University, 251 Mercer Street, New York, NY 10012, USA.}
\author{Noah Ziethen}%

\address{%
 Max Planck Institute for Dynamics and Self-Organization, Am Fassberg 17, 37077 G\"{o}ttingen, Germany
}
\author{Karen Alim}
\ead{Corresponding author E-mail: k.alim@tum.de}
\address{%
Center for Protein Assemblies (CPA) and Department of Bioscience, School of Natural Sciences, Technical University of Munich, Ernst-Otto-Fischer-Str. 8, 85748 Garching, Germany
}%
\address{%
 Max Planck Institute for Dynamics and Self-Organization, Am Fassberg 17, 37077 G\"{o}ttingen, Germany
 }
\vspace{10pt}

\begin{abstract}
Understanding vascular adaptation, namely what drives veins to shrink or grow, is key for the self-organization of flow networks and their optimization. From the top-down principle of minimizing flow dissipation at a fixed metabolic cost within flow networks, flow shear rate resulting from the flows pervading veins is hypothesized to drive vein adaptation. Yet, there is no bottom-up derivation of how flow forces impact vein dynamics. From the physical principle of force balance, shear rate acts parallel to vein walls, and hence, naively shear rate could only stretch veins and not dilate or shrink them. We, here, resolve this paradox by theoretically investigating force balance on a vein wall in the context of the vascular network of the model organism \pp. We propose, based on previous mechanical studies of cross-linked gels, that shear induces a nonlinear, anisotropic response of the actomyosin gel, making up vein walls that can indeed drive vein dilatation. Furthermore, our force balance approach allows us to identify that shear feedback occurs with a typical timescale and with a typical target shear rate that are not universal properties of the material but instead depend smoothly on the location of the vein within the network. In particular, the target shear rate is related to the vein's hydrostatic pressure, which highlights the role of pressure in vascular adaptation. Finally, since our derivation is based on force balance and fluid mechanics, we believe our approach can be extended to vascular adaptation in other organisms.
\end{abstract}

%
%
%
%
%

\section{Introduction}

\begin{figure}[h!]
\centering
\includegraphics[width=0.6\textwidth]{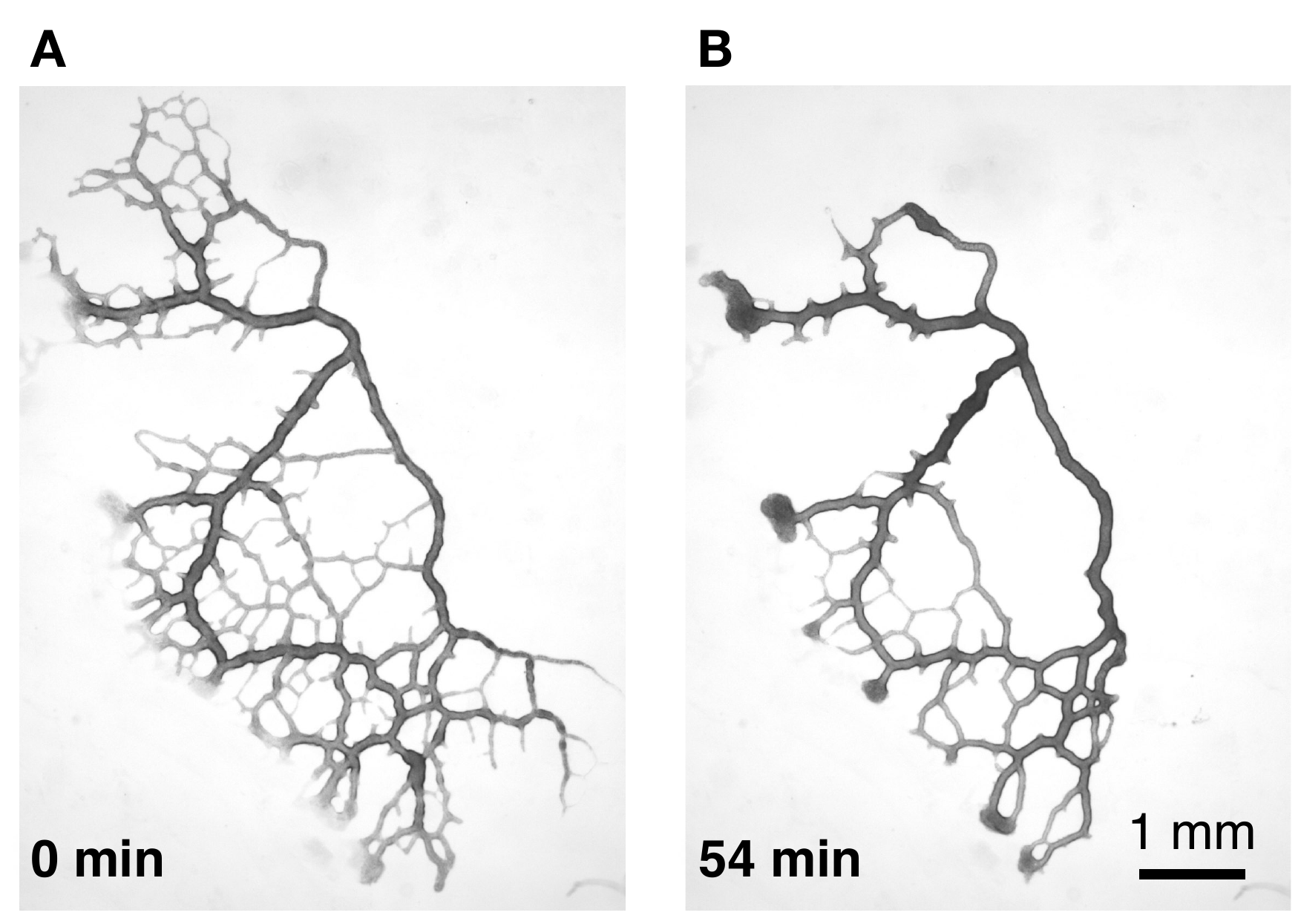}
\caption{Bright-field images of a reorganizing specimen of \ppf, from a reference state (A) and $54~\mathrm{min}$ later (B). Adapted with permission from Ref.~\cite{marbach2021network}.}
\label{fig:intro}
\end{figure}

Vascular flow networks continuously reorganize by growing new veins or shrinking old ones~\cite{lucitti2007vascular,chen2012haemodynamics,hu2013adaptation}, to optimize specific functions such as nutrient or information distribution or to adapt to changing environmental cues. As an example, we show in Fig.~\ref{fig:intro} the spontaneous reorganization of the slime mold \ppf~over the course of a few hours, which shows significant vein trimming. Vascular adaptation is seen across the plant and animal realms: from blood vasculature~\cite{kurz2000physiology,hove2003intracardiac,chen2012haemodynamics,zhou1999design}, via leaf venation in plants~\cite{corson2009silico,ronellenfitsch2016global} to vein networks making up fungi and slime molds~\cite{tero2010rules,Alim2013}. Understanding vascular adaptation is crucial to probe healthy development~\cite{chen2012haemodynamics} and disease growth~\cite{Meyer:2008,pries2009structural}. 

At steady state, Murray's law~\cite{murray1926physiological} provides a rather reliable prediction of network morphologies across a variety of animals and plants~\cite{west1997general,kassab2006scaling,west1997general,mcculloh2003water,akita2016experimental,fricker2017automated}. Briefly, we recapitulate the main phenomenological ingredients yielding this law, based on the principle of minimum work. Murray stipulated~\cite{murray1926physiological} that energy dissipation in a vein of radius $a$ and length $L$ is given by shear dissipation and metabolic expense to sustain the vein
\begin{align}
\mathcal{E}= \frac{1}{2} \frac{Q^2}{R} + \pi  b L a^2 = \frac{4 \mu L Q^{2}}{\pi a^{4}}+ \pi b L a^{2}.
\label{eq:Murray_energy}
\end{align}
where $R = \pi a^4 / 8 \mu L$ is the vein's hydraulic resistance assuming Poiseuille flow in the vein, $b$ is a local metabolic constant per unit volume, $Q$ the flow rate, and $\mu$ the fluid's viscosity. The principle of minimum energy expense suggests searching for the minimum of $\mathcal{E}$ with respect to the vein radius $a$, which gives the relation $a^{6}=\frac{8 Q^{2} \mu}{b \pi^{2}}$. If we calculate the shear rate $\tau =\frac{4 Q}{\pi a^{3}}$ in this optimized state, we obtain that shear rate is constant and equal to an optimal value $\tau =\sqrt{b/\mu} \equiv \tau_0$, independent of vein radius $a$. 

Beyond the steady state, dynamic adaptation of veins has often been modeled relying on the following \textit{phenomenological} adaptation equation for individual veins:
\begin{equation}
\frac{1}{a} \frac{da}{dt} = \frac{1}{t_{\rm adapt}} \left( f(\tau) - f(\tau_0)\right)
\label{eq:phenomenological}
\end{equation}
where $a(t)$ is the vein radius at time $t$, $t_{\rm adapt}$ an adaptation time scale, $\tau$ the local shear rate in the vein and $f(\tau)$ is a monotonically increasing function such that $f(\tau = 0) =0$~\cite{tero2007mathematical}. 
We will specify $f$'s specific form later. Note that we will, here, use $\tau$ to represent the shear \textit{rate} in a vein. It is also common to discuss shear \textit{stress} $\sigma$, which is simply related to the shear rate as $\sigma = \mu \tau$ where $\mu$ is again the fluid's viscosity. At steady state, the shear rate is constant and equal to the target shear rate value $\tau = \tau_0$, consistent with Murray's law. The parameters $t_{\rm adapt}$ and $\tau_0$ are usually taken to be network dependent but constant across the organism~\cite{Taber1998Model,hacking1996shear,hu2012blood,tero2007mathematical,akita2016experimental, baumgarten2013functional,ronellenfitsch2016global,pries1998structural,pries2005remodeling,secomb2013angiogenesis,hu2013adaptation,bonifaci2017revised,bonifaci2012physarum}.

The variety of functions $f(x)$ used in such \textit{phenomenological} models already points to a lack of consensus. Some works investigate $f(\tau) \sim |\tau|$~\cite{Taber1998Model,hacking1996shear,hu2012blood,tero2007mathematical,akita2016experimental, baumgarten2013functional,bonifaci2017revised,bonifaci2012physarum} with possible generalizations and extensions~\cite{ronellenfitsch2016global}, while others consider $f(\tau) \sim \log(\tau)$~\cite{pries1998structural,pries2005remodeling} and extensions~\cite{secomb2013angiogenesis}. Notably, a rather recent work~\cite{hu2013adaptation} follows Murray's law of minimizing energy dissipation to arrive at $f(\tau) \sim \tau^2$. However, there is currently no effort to understand the origin of such an adaptation rule bottom-up. Therefore we lack the chance to validate the functional dependence on shear rate. 

In addition, an adaptation rule with shear rate driving tube dilation or shrinkage is rather counter-intuitive from a mechanical perspective. In all vascular biological networks~\cite{Taber1998Model,hacking1996shear,hu2012blood,tero2007mathematical,akita2016experimental, baumgarten2013functional,ronellenfitsch2016global,pries1998structural,pries2005remodeling,secomb2013angiogenesis,hu2013adaptation}, the same laws govern laminar flow through slender veins: the shear rate $\tau$, evaluated on the surface of a vein, acts on the longitudinal direction along the vein. Therefore, naively shear rate can only \textit{extend a vein longitudinally}, but dilatation or shrinking, namely changes in the vein radius $a(t)$, may not arise. 

In this work, we reconcile the paradox of how shear rate can drive vein radius changes and derive Eq.~\eqref{eq:phenomenological} by establishing a detailed mechanical force balance on a vein wall. To this end, we focus our derivation on the broadly studied model organism \ppf. One important specificity of \pp~is that its veins are encapsulated in an actomyosin fiber cortex. For the latter, recent experimental studies~\cite{gardel2008mechanical,janmey2007negative} show that cross-linked actin fibers respond anisotropically to shear and, hence, may dilate or shrink veins instead of just acting in the longitudinal direction, as discussed in detail by us. The detailed characteristics of the actomyosin fiber cortex determine the exact shape of $f(\tau)$, which should be increasing with $\tau$ and for \pp~is well approached by $f(\tau) \propto \tau^2$. This investigation also shows that $f(\tau)$ could take other shapes according to the cortex's mechanical properties that vary across biological systems. Finally, and in contrast with previous assumptions, we find that $t_{\rm adapt}$ and $\tau_0$ are \textit{local} quantities that vary smoothly and slowly throughout the network but are location-specific. Interestingly, we find that $\tau_0$ is related to local hydrostatic pressure, which confirms the role of pressure in vascular adaptation. Our work, therefore, opens up the possibility to investigate vascular reorganization; see our accompanying mostly experimental work~\cite{marbach2021network}.

\section{Model setup}

\subsection{Low Reynolds number flows in a contractile network}

\begin{figure}[h!]
\centering
\includegraphics[width=0.85\textwidth]{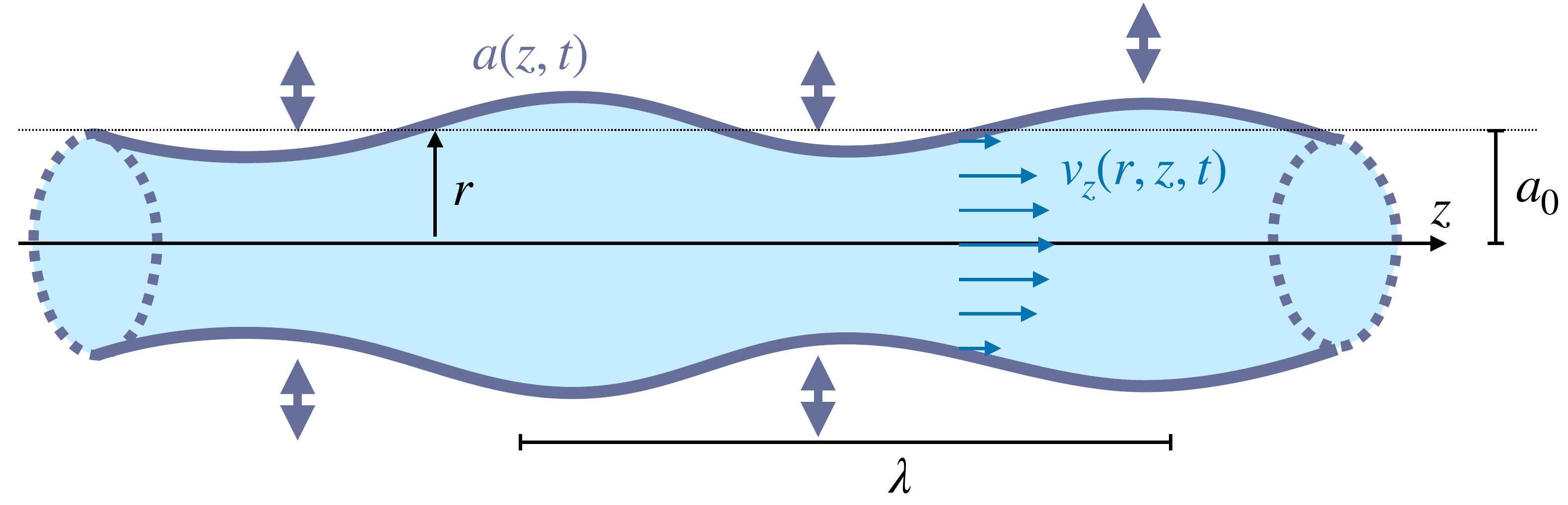}
\caption{Sketch of a vein of radius $a(z,t)$ filled with fluid flowing with longitudinal speed $v_z(r,z,t)$. }
\label{fig:sketch}
\end{figure}

We model flow in a single vein filled with cytoplasmic fluid considered incompressible (see Fig.~\ref{fig:sketch}). The vein radius undergoing rhythmic, peristaltic contractions is given by $a(z,t)$, where $z$ is the longitudinal coordinate along the vein and $t$ time. The radial coordinate is denoted by $r$. We consider for simplicity that the vein has cylindrical symmetry. The flow field inside the vein is $v_r(r,z,t)$ in the radial direction and $v_z(r,z,t)$ along the vein axis, while pressure is written as $p(r,z,t)$.

We start by calculating relevant non-dimensional numbers characterizing the flow to simplify the Navier-Stokes equations describing the flow field in the contractile vein. 

First, vein contractions can be treated within the lubrication approximation, where the wavelength of the peristaltic contractions $\lambda$ is larger than the average vein radius $a_0$. In fact, in \pp, the contractile wavelength typically extends over the organism's size~\cite{Alim2013}, $ \lambda \simeq 5 \mathrm{mm}$, while the 
 typical radius of a vein is $a_0 \sim 100 \mathrm{\mu m}$. Hence, we can define the small non-dimensional number $\epsilon_{\lambda} = a_0/\lambda \ll 1$. We can, therefore, apply the lubrication approximation, which will allow us to keep only first-order terms in $\epsilon_{\lambda}$ in the Navier-Stokes equations using the hierarchy of derivatives  $ \frac{\partial v_z}{\partial r} \sim \frac{1}{\epsilon_{\lambda} } \frac{\partial v_r}{\partial r} \sim \frac{1}{\epsilon_{\lambda} }  \frac{\partial v_z}{\partial z} \sim \frac{1}{\epsilon_{\lambda}^2} \frac{\partial v_r}{\partial z} $, see also \textit{e.g.} Ref.~\cite{grun2006thin,marbach2018transport}. 

Second, the appropriate Reynolds number for flows in contractile veins is $Re = \frac{\rho c a_0}{\mu} \epsilon_{\lambda} $~\cite{li1993non} where $c$ is the typical speed of the contractile wave, $\mu$ is the dynamic viscosity and $\rho$ the fluid density. 
A typical value for the cytoplasmic viscosity is $\mu = 1.5 \times 10^{-3}~\mathrm{Pa\cdot s}$~\cite{swaminathan1997photobleaching,puchkov2013intracellular}. 
The speed of the contractile wave is $c = \lambda/T$ where $ \lambda \simeq 5 \mathrm{mm}$ is the scale of the organism and $T \simeq 1-2 \mathrm{min}$ is the contraction period measured in \pp~\cite{stewart1959protoplasmic,isenberg1976transformation,Alim2013}. We find $c \simeq 50 \mathrm{\mu m/s}$ also agreeing with flow velocities inside veins, directly measured from our own velocimetry measurements~\cite{marbach2021network}. Taking the density $\rho$ of water, this yields $Re \simeq 0.003 \epsilon_{\lambda}$ where $\epsilon_{\lambda} \ll 1$, which enables us to neglect non-linear terms in the Navier-Stokes equations. 

Finally, we evaluate the Womersley number $\alpha_W = \sqrt{\frac{\rho a_0^2 \omega}{\mu}}$~\cite{womersley1955xxiv} with $\omega = 2\pi/T$. The Womersley number quantifies the relative importance of time-dependent inertial terms in the Navier-Stokes equation with respect to viscosity terms. We find $\alpha_W \simeq 0.02$.  We can therefore neglect time-dependent inertial terms in the Navier-Stokes equations. 

All in all, the Navier-Stokes equations describing the flow field inside the vein reduce to 
\begin{equation}
\begin{cases}
    \displaystyle 0 = - \frac{\partial p}{\partial z} + \mu \frac{1}{r} \frac{\partial}{\partial r} \left( r  \frac{\partial v_z}{\partial r}  \right),  \\
   \displaystyle \frac{1}{r}\frac{\partial (r v_r)}{\partial r} + \frac{\partial v_z}{\partial z}  = 0.
\end{cases}
\end{equation}
Notice that although we arrived at such equations by determining that the non-dimensional numbers $\epsilon_{\lambda}$, $Re$ and $\alpha_W$ are small in \pp, these numbers are usually small across a wide variety of biological systems~\cite{marbach2018transport}. 

Solving these equations in the limit of the lubrication approximation where the dependence of $\frac{\partial p}{\partial z}$ on $r$ can be neglected yields the flow profiles~\cite{Alim2013}
\begin{equation}
    \begin{cases}
    \displaystyle v_z(r,z,t) = &-\frac{a^2(z,t)}{4 \mu} \frac{\partial p}{\partial z} \left( 1 - \left( \frac{r}{a(z,t)}\right)^2 \right), \\
    \displaystyle  v_r(r,z,t) = &\frac{\partial a(z,t)}{\partial t} \frac{r}{a(z,t) }\left( 2 - \left( \frac{r}{a(z,t)}\right)^2 \right) \\
    &- \frac{a(z,t)^2}{4 \mu} \frac{\partial p}{\partial z} \frac{r}{a(z,t) } \frac{\partial a(z,t)}{\partial z}\left( 1 - \left( \frac{r}{a(z,t)}\right)^2 \right).
    \end{cases}
    \label{eq:flows}
\end{equation}
Finally, conservation of mass imposes that, along the vein,
\begin{equation}
    \frac{\partial }{\partial t} ( \pi a^2(z,t))  = - \frac{\partial }{\partial z} \left(2 \pi \int_0^{a(z,t)} u_z(r,z,t)  r dr \right) = - \frac{\partial }{\partial z} \left( - \frac{\pi a^4}{8 \mu} \frac{\partial p}{ \partial z}\right)
    \label{eq:massConservation}
\end{equation}
and allows us to infer an equation relating the pressure field $p$ to the contraction profile $a(z,t)$. The remaining missing equation to fully characterize the four variables $p$, $a(z,t)$, $v_r$, and $v_z$ corresponds to the force balance on the vein wall, which we detail below. 

\subsection{Elastic and plastic adaptation of the vein radius $a$}

We will now seek an equation on the evolution in time of the vein radius $a(z,t)$, ultimately via force balance on the vein wall. To understand precisely what this equation will mean, it is relevant to discuss first a few physical characteristics of the time variations of $a$. 

Firstly, the radius of the vein evolves in time according to different processes. Some deformations are \textit{elastic} and short lived, while other deformations occur on longer timescales and correspond to growth or disassembly of the wall material, they are \textit{plastic} deformations. To distinguish these two timescales or two kinds of deformations, we consider the averaging operator $\langle \cdot \rangle$, which is an average over short timescales, typically corresponding to elastic deformations and to the short peristaltic contractions in \pp. $\langle a \rangle(t)$ therefore only has variations over long timescales, corresponding to growth or disassembly. In contrast, $(a(z,t) - \langle a \rangle(z,t))$ corresponds to short timescale elastic deformations. 

Secondly, longitudinal variations along the vein segment axis $z$, $\partial a/\partial z$ can be neglected, as we assumed $\epsilon_{\lambda} \ll 1$ and $\partial a/\partial z \sim \epsilon_{\lambda}$.

Finally, we must also discuss how the vein wall's thickness evolves over time. Direct observations from organisms analyzed in our companion work~\cite{marbach2021network} suggest that the ratio of the thickness of the vein wall $e$ to the vein size $a$ is roughly constant throughout experiments, $e/a \simeq \text{const.}$ While a detailed analysis of vein thickness in \pp~is beyond this work's scope, we can infer the typical ``eye-rule'' $e \simeq 0.1 a$. Note that this ratio remains quite accurate also for shrinking veins because veins do not shrink to $a = 0$; rather, they shrink to some small value $a \simeq 10-20~\mathrm{\mu m}$ and then are ``retracted'' into the network. Hence, the proportionality relation is approximately satisfied even in small veins, $e \simeq 0.1 a$. In the following, we will, thus, only model the dynamics for $a(t)$, assuming that the dynamics for the vein thickness closely follow that of the radius $a(t)$. 

\section{Force balance on a vein wall}

We will now balance forces on a vein wall for a small, ring-like vein segment of infinitesimal length $\dl$ and radius $a$. Since we are interested in vein adaptation dynamics, that is to say, in the time evolution of $a$, we focus on radial forces only since these are the only ones that may contribute to radius dilatation or shrinkage. We, thus, enumerate the different forces at play on the vein wall segment of length $\dl$ in the radial direction. 

\subsection{Hydrodynamic forces.}

 Let $\Pi = - p \, \mathbb{I} + \sigma$ be the tensor characterizing hydrodynamic forces per unit area, where $\sigma$ is the hydrodynamic stress tensor and $\mathbb{I}$ is the identity matrix. The hydrodynamic forces acting in the radial direction on the vein's wall are then
 \begin{equation}
     \df_{\rm hydro} =  \delta S \, \bm{e_r} \cdot (\Pi - p_{\rm ext} \mathbb{I}) \cdot \bm{e_r} = \delta S  (\Pi_{rr} - p_{\rm ext} )
 \end{equation}
 where $\bm{e_r}$ is the unit vector in the outward radial direction, $\delta S = 2\pi a \dl$ is the infinitesimal surface area of the vein, and $p_{\rm ext}$ is the atmospheric pressure exerted uniformly across the organism on the outer side of vein walls.
 
 To calculate radial forces on the vein's wall, we will thus need the radial components of the tensor $\Pi$,
    \begin{equation}
        \begin{cases}
        \displaystyle \Pi_{rr}|_{r=a} = - p + 2 \mu \frac{\partial v_r}{\partial r}, \\
        \displaystyle \Pi_{rz}|_{r=a} = \mu \left( \frac{\partial v_r}{\partial z} + \frac{\partial v_z}{\partial r} \right)
        \end{cases}
    \end{equation}
    where we also calculate $\Pi_{rz}|_{r=a}$ since we will use it later. 
    Within the lubrication approximation, we have the hierarchy $ \frac{\partial v_z}{\partial r} \sim \frac{1}{\epsilon } \frac{\partial v_r}{\partial r} \sim \frac{1}{\epsilon }  \frac{\partial v_z}{\partial z} \sim \frac{1}{\epsilon^2} \frac{\partial v_r}{\partial z} $ and further $p \sim \mu \frac{1}{\epsilon^2} \frac{\partial v_r}{\partial r}$. Keeping only highest order terms in $\epsilon$ yields
    \begin{equation}
        \begin{cases}
        \displaystyle \Pi_{rr}|_{r=a} \simeq - p  \\
        \displaystyle \Pi_{rz}|_{r=a} \simeq - \frac{4 \mu \overline{v}}{a}  
        \end{cases}
    \end{equation}
    where $\overline{v}(z,t) = - \frac{a^2(t)}{8 \mu} \frac{\partial p}{\partial z}$ is the cross-sectional average of $v_z$. 
The $\Pi_{rz}$ component of the stress corresponds to the shear stress, which is related to the shear \textit{rate} $\tau = \frac{4 \bar{v}}{a} =  \frac{4 Q}{\pi a^3}$ where $Q = \pi a^2 \bar{v}$ via the dynamic viscosity: $|\Pi_{rz}|_{r=a}| = \mu |\tau|$. 

We obtain that the resulting hydrodynamic forces on the vein's wall are simply related to the pressure imbalance, with larger pressures differences between the hydrostatic pressure $p$ and the atmospheric pressure $p_{\rm{ext}}$ dilating veins as expected
 \begin{equation}
     \df_{\rm hydro} =  (p - p_{\rm ext} )2\pi a \dl.
     \label{eq:Fhydro}
 \end{equation}
 Importantly, we remark that shear stress has no contribution to the hydrodynamic radial forces. 
    
\subsection{Potential forces.}
    Let $\mathcal{H}$ be the Hamiltonian of the membrane wall, describing the potential energy related to all conservative forces in the system. These correspond to elasticity, bending, surface tension, stretching forces, \textit{etc}. The corresponding stress, or force per unit surface, is $\sigma_{\rm circum} = - \frac{1}{\delta S} \frac{\delta \mathcal{H}}{\delta a}$, where $\delta S = 2 \pi a \dl $ is the infinitesimal surface area of the vein segment. For vein walls such as the ones making up \pp, the Hamiltonian can assume various forms~\cite{olufsen1999structured,storm2005nonlinear,barthes2016motion,alim2017mechanism,julien2018oscillatory}. Here, we assume Hookean-type feedback as a zeroth order model for elasticity only. Note that the following derivation may be done in a similar way for other Hamiltonians, including \textit{e.g.,~}bending terms. In practice, in the limit of small deformations, these should yield similar contributions to $\sigma_{\rm wall}$. With elastic terms only, the potential force in the radial direction on the vein wall sums up to the so-called circumferential stress (see Eq.~(2.4) of Ref.~\cite{takagi2011peristaltic}) multiplied by the vein's surface area
    \begin{equation}
       \df_{\rm circum} = \sigma_{\rm circum} \delta S = - \frac{E}{e(1-\nu^2)} (a(t) - a_{0}(t)) 2 \pi a \dl
       \label{eq:elastic}
    \end{equation}
    where $E$ is Young's modulus, in $\SI{}{Pa}$, characterizing the wall's elasticity, $\nu$ is the material's Poisson's ratio, which is a number with no units, $e$ is the thickness of the wall, which is assumed to be a thin elastic shell and $a_{0}$ is a reference radius.
    Note that $a_{0}$ may slowly evolve with space and time as the vein slowly adapts. In fact as $a(t) - a_{0}(t)$ characterizes the vein's \textit{elastic} deformation, we may assume that $a_{0}(t)$ closely follows the short timescale average of $a$, namely $\langle a \rangle(t)$, such that $\langle a(t) - a_{0}(t) \rangle \simeq 0$. Finally, recall that $a \simeq 0.1 e$.
    As $\sigma_{\rm circum}$ scales as $a/e$ according to Eq.~\eqref{eq:elastic}, but $e\simeq 0.1 a$, we may thus infer that $\sigma_{\rm circum}$ is mostly \textit{independent} of the vein radius $a$. 
    
\subsection{Active forces.} 
    We denote $\sigma_{\rm active}$ the active stress operated by the actomyosin cortex~\cite{radszuweit2013intracellular,alonso2017mechanochemical}. We assume these active stresses induce short timescale dynamics that are purely elastic deformations. The short time scale here corresponds to the period of the peristaltic contractions. 
    A chemical potentially triggers these active forces driving contractions within the cytoplasm~\cite{radszuweit2013intracellular,alonso2017mechanochemical,alim2017mechanism,julien2018oscillatory}. Here, we assume that the chemicals are initially well mixed, meaning that the organism has received no localized food or chemical stimulus spatially altering the chemical balance. Throughout the analysis, we further assume that these compounds remain well mixed in the absence of external stimuli. This allows us to consider in particular that the active forces do not show any significant trend on long timescales, such that 
    \begin{equation}
        \langle \sigma_{\rm active} \rangle  \simeq \text{const.}
    \end{equation}
    where $\langle . \rangle$ denotes the average over short timescales. We will be more specific as to what enters that constant later. The radial force resulting from this active stress is simply
    \begin{equation}
        \df_{\rm active} =  \sigma_{\rm active} \delta S.
        \label{eq:Factive}
    \end{equation}

    
\subsection{Nonlinear anisotropic feedback forces from shear stress.}

Shear stress, here denoted as $\Pi_{rz} = \sigma_{rz}$, is considered to be the dominating mechanical force for growth induced by shearing cells in numerous experiments~\cite{hoefer2013biomechanical,koller1993role}. The importance of shear stress forces for adaptation dynamics could be explained by mechanosensitive pathways~\cite{fernandes2018hemodynamic} or other chemical pathways that regulate the dilatation of veins~\cite{lu2011role, godbole2009nadph}. However, when interested in force balance, shear stress exerts a force in the \textit{longitudinal} direction on the vein wall and, hence, can not contribute \textit{a priori} to the radial forces that dilate or contract the vein. 

However, shearing the actomyosin cortex can lead to a significant \textit{anisotropic} response, namely to negative \textit{normal} stress, \textit{i.e.} outward, because of the viscoelastic-like nature of the actomyosin gel~\cite{gardel2008mechanical,janmey2007negative,vahabi2018normal} as detailed below. 
Overall this means that we need to add a radial extensional force 
\begin{equation}
\label{eq:fshear}
    \df_{\rm gel}= \sigma_r(\sigma_{rz}) \delta S
\end{equation}
where $\sigma_{rz} = \Pi_{rz}|_{r=a}$ is the shear stress at the wall and $\sigma_r(\sigma)$ has units of a stress and depends on shear stress at the wall, possibly in a nonlinear way. This anisotropic response does act in the radial direction and, hence, solves the apparent paradox of how shear stress, via force balance, can act in the radial direction. It is one of the key steps in our derivation.

\begin{figure}
\centering
\includegraphics[width=0.9\textwidth]{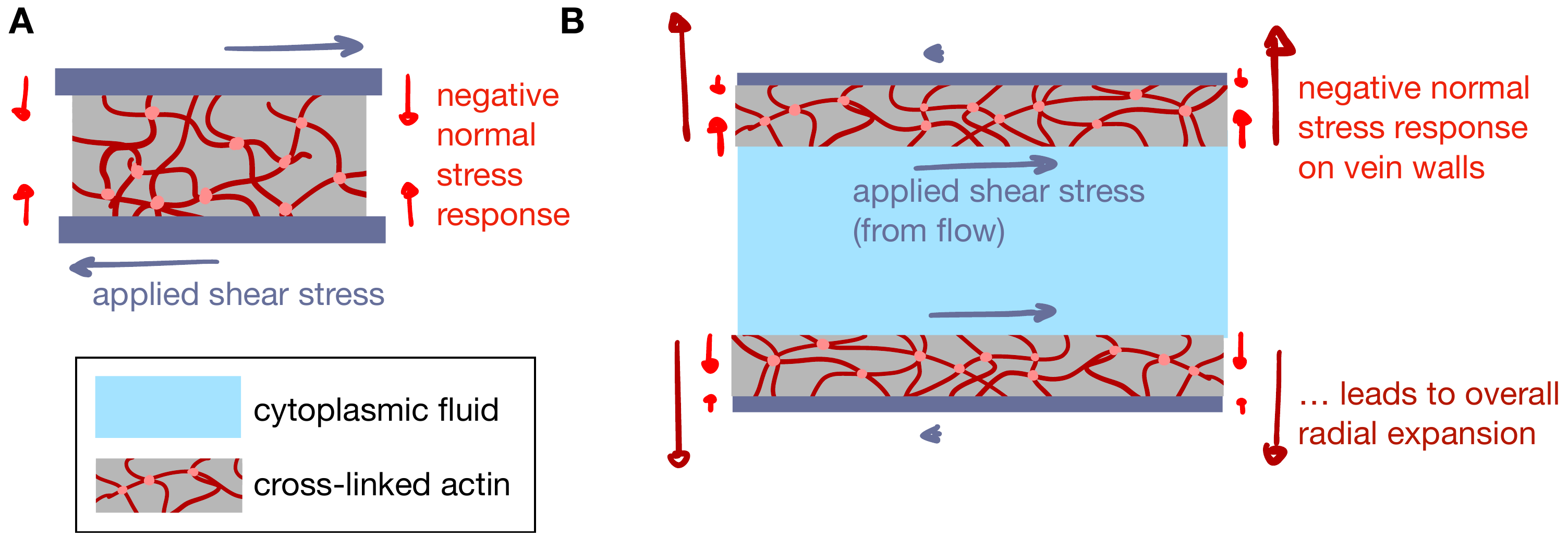}
\caption{Shear stress leads to radial expansion. (A) Experimental setting of Ref.~\cite{janmey2007negative} where two plates (dark blue) are shearing a crosslinked gel and a negative normal stress response is observed that tends to bring the plates together. (B) Analogous effect in the vein geometry, cut along the  longitudinal direction of a vein, where the gel makes up the vein walls. Shear stress from cytoplasmic streaming results in negative compression of the outer actin gel, which leads to an expansion of the vein radius.}
\label{fig:Janmey}
\end{figure}

How can we understand this anisotropic gel response?
Ref.~\cite{janmey2007negative} explores the normal stress response in a variety of biogels that are crosslinked through semi-flexible or rigid filaments. The gels are sheared between two plates, and experiments identify an orthogonal response with a ``negative'' normal stress, meaning that the plates are attracted towards one another -- see Fig.~\ref{fig:Janmey}-A. This effect is due to the rigid filaments crosslinking the gels that resist shearing by bringing fibers closer to one another, creating the normal stress response~\cite{janmey2007negative,gardel2008mechanical}. This negative normal anisotropic response is seen across many forms of biogels~\cite{kang2009nonlinear,vahabi2018normal} and is confirmed by minimal numerical models~\cite{conti2009cross} and nonlinear elastic theory~\cite{vahabi2018normal}. 

Does the anisotropic gel response apply in the case of the \pp~cortex?
Ref.~\cite{janmey2007negative} performed measurements on non-active interconnected actin filaments but also on various other crosslinked filaments. Refs.~\cite{janmey2007negative,gardel2008mechanical} have further identified that the anisotropic response is likely due to the material's rigid crosslinks. Crosslinks, including passive crosslinks, are still present in the active actomyosin gel; hence, it is a valid assumption that such a negative normal response would be maintained in gels with an active component such as the one studied here.

%
%
%

How does the anisotropic gel response translate to a vascular geometry?
In our vascular setting with fluid enclosed by an actin cortex gel, the vein wall is sheared between the fluid-gel and the air-gel boundaries; see Fig.~\ref{fig:Janmey}-B. Hence, the fluid-gel and air-gel boundaries tend to move toward one another. Because there is little resistance on the air-gel side, the gel can relax stress by expanding further on the free-air side. Effectively this means that the fluid-gel boundary moves away from the fluidic region, and, hence, the fluid-gel boundary expands; in other words, the vein radius increases. Since the air-gel boundary is free, this also means that the wall thickness does not change significantly, and the approximation $e\simeq 0.1 a$ is consistent. In conclusion, the radial stress expands the vein's radius and, therefore, $\sigma_r(\sigma) > 0$. Note that the vein's fluid and wall mass are not necessarily conserved during the adaptation process, but this is not contradictory. Indeed, fluid mass is brought in and out from other parts of the network, and wall structure is continuously deformed over long timescales, corresponding to \textit{e.g., ~}actin fiber turnover and rearrangements~\cite{salbreux2012actin,fischer2016rheology}.

What is the functional dependence of the normal stress on the shear stress?
Nonlinear elasticity theory can predict the anisotropic normal stress due to shear stress~\cite{conti2009cross,horgan2011normal,unterberger2013new,holzapfel2014affine,horgan2017poynting,vahabi2018normal}.
The magnitude of the normal stress response depends on the mechanical properties of the gel and of the crosslinking filaments. Furthermore, the normal response scales linearly or quadratically with the shear rate~\cite{conti2009cross,holzapfel2014affine,vahabi2018normal}. Quadratic models accurately reproduce the negative normal stresses observed in Ref.~\cite{janmey2007negative}. We, therefore, assume in the following that the nonlinear anisotropic stress response can be written 
\begin{equation}
\sigma_r(\sigma) = \frac{1}{\sigma_c}\sigma^2,
\end{equation}
where $\sigma_c > 0$ is a characteristic shear stress quantifying the responsiveness of the gel wall material. Other functional forms, that are monotonically increasing with $\sigma > 0$ and that verify $\sigma_r(\sigma = 0) = 0$, such as $\sigma_r(\sigma) = |\sigma|$ would not change the main results of this work. Here, only the anisotropic response is essential to derive an adaptation rule from force balance. 

Finally, we need to specify for the anisotropic response what timescales are involved, precisely, whether $\sigma_{rz}$ in Eq.~\eqref{eq:fshear} corresponds to short, elastic, or long, plastic, timescale contributions. In \pp, we observe in experiments that vein adaptation, compared to the input shear signal, happens with a time delay~\cite{marbach2021network}. This time delay is typically at least of the order of a contraction period. Hence, the time delay washes out short timescale stress contributions, and only the long timescale dependence of shear is relevant in Eq.~\eqref{eq:fshear}. More generally, we may expect the gel's anisotropic response to occur with a timescale corresponding to actin fiber rearrangements. Therefore the anisotropic response varies only over long timescales.  
In the following, we prefer to express equations in terms of the shear rate $\tau = \sigma_{rz}/\mu$ such that we may rewrite the radial anisotropic gel response as 
\begin{equation}
    \df_{\rm gel}  = \sigma_r(\mu \langle \tau\rangle) \delta S  = \mu \frac{\langle\tau \rangle^2}{\tau_c} 2\pi a \dl,
    \label{eq:fgel}
\end{equation}
where we included the fact that the relevant shear is over long timescales. 
 We also define $\tau_c = \sigma_c/\mu$ as the associated characteristic shear rate.

%
%

\subsection{Friction forces.}    
    Finally, it is natural to expect that the actomyosin cortex and outer membrane layer will resist strong plastic deformations in vasculature morphology. We expect this to be due to inherently long timescales, corresponding to \textit{e.g., ~}actin fiber turnover and rearrangements~\cite{salbreux2012actin,fischer2016rheology}. Because actin fiber turnover happens on a much longer timescale than myosin turnover~\cite{salbreux2012actin,fischer2016rheology}, we expect such timescales to be much longer than elastic timescales or the contraction period. To introduce this long timescale response in our equations, we add a friction force. We write the friction force on an infinitely small vein ring as
    \begin{equation}
        \df_{\rm friction} = - \gamma(a,e) \frac{d \langle a \rangle}{d t} \dl
        \label{eq:Ffriction}
    \end{equation}
    where $\gamma(a,e)$ is a line friction coefficient that models these long adaptation phases. Here, we assumed a friction force linear in the long-time deformation speed $\frac{d\langle a \rangle}{dt}$ since (a) only the long-time deformation corresponds to plastic deformations and (b) the linearity is expected for typical slow, low Reynolds number, motion~\cite{happel2012low}. It is not obvious what the form of $\gamma(a,e)$ should be, and to simplify the derivations, we assume in further calculations that it is constant $\gamma(a,e) = \gamma$. 
Typically $\gamma$ depends on the mechanical properties of the gel cortex.

\subsection{Force balance on a vein segment}

Gathering all forces as outlined above, we can now write the balance of forces acting on a vein section of length $\dl$ as
\begin{equation}
    \delta m \frac{d^2 a}{dt^2} = \df_{\rm hydro} +  \df_{\rm cricum} + \df_{\rm active} + \df_{\rm gel} + \df_{\rm friction}    
\end{equation}
where $\delta m = 2 \pi a e \rho_{C} \dl$ is the infinitesimal mass of the vein wall section and $\rho_{C}$ is the density of the cytoskeletal wall. Using the expressions derived above for the forces, Eq.~\eqref{eq:Fhydro}, Eq.~\eqref{eq:elastic}, Eq.~\eqref{eq:Factive}, Eq.~\eqref{eq:fgel} and Eq.~\eqref{eq:Ffriction}, we obtain
\begin{equation}
    2 \pi a e L \rho_{C} \frac{d^2 a}{dt^2} = \left((p - p_{\rm ext})  - \frac{E}{1-\nu^2} \frac{\left( a - \langle a \rangle \right)}{e}  + \sigma_{\rm active} + \mu \frac{ \langle \tau \rangle^2}{\tau_c} \right) 2 \pi a L - \gamma L \frac{d\langle a\rangle}{dt}.
    \label{eq:forcebalanceA}
\end{equation}
Note that we neglected any fluctuating forces in this simple force balance, especially as they would eventually be averaged out in the long timescale we focus on here.

\section{Long time adaptation model}
We now aim to simplify the force balance Eq.~\eqref{eq:forcebalanceA}. First, we make a standard overdamped approximation and neglect inertial terms since the time scales associated with the relaxation of inertia are much faster than any other time scales in the system.~\footnote{In fact, we have $\rho_{C} \simeq 10^3~\mathrm{kg/m^3}$ and $\gamma \simeq 6 \pi \mu$. We seek an upper bound on the relaxation of inertia by using a viscosity $\mu \simeq 10^{-3}~\mathrm{Pa.s}$ close to that of water, while we expect much higher viscosity of the cortex. This implies that the time scale for the relaxation of inertia is at most $t_{\rm inertia} = \frac{2 \pi a e\rho_C}{\gamma} \simeq \frac{\rho_C  a e}{3 \mu} \simeq \frac{10^3 \times 50 \times 10^{-6} \times 5 10^{-6}}{3\times 10^{-3}} \simeq 10^{-4} s$. Hence, $t_{\rm inertia}$ is much shorter than any other relevant time scale in the system.} We obtain that Eq.~\eqref{eq:forcebalanceA} simplifies to
\begin{equation}
    \frac{d\langle a\rangle}{dt} = \frac{2 \pi a}{\gamma} \left((p - p_{\rm ext})  - \frac{E}{1-\nu^2} \frac{\left( a - \langle a \rangle \right)}{e}  + \sigma_{\rm active} + \mu \frac{ \langle \tau \rangle^2}{\tau_c} \right) 
    \label{eq:forcebalance2}
\end{equation}
We now use our timescale separation assumption: some variables demonstrate either long timescale dynamics or short timescale dynamics in line with radius dynamics that have both short and long timescales. Long timescale dependencies will, from here on, be written as $\langle X \rangle(t)$ for any variable $X$.  

\subsection{Short timescales}

We first focus on short timescales. 
To observe short timescale dynamics, we can take the short-time average of Eq.~\eqref{eq:forcebalance2} and subtract it back from Eq.~\eqref{eq:forcebalance2}. We obtain
\begin{equation}
\begin{split}
    0 = &\frac{2 \pi a}{\gamma} \left( (p - p_{\rm ext})  - \frac{E}{1-\nu^2} \frac{\left( a - \langle a \rangle \right)}{e}  + \sigma_{\rm active} + \mu \frac{ \langle \tau \rangle^2}{\tau_c} \right) \\
    &- \bigg\langle \frac{2 \pi  a }{\gamma} \left((p - p_{\rm ext}) - \frac{E}{1-\nu^2} \frac{\left( a - \langle a \rangle \right)}{e}  + \sigma_{\rm active} + \mu \frac{ \langle \tau \rangle^2}{\tau_c} \right) \bigg\rangle.
    \label{eq:forcebalance31}
\end{split}
\end{equation}
We can simplify the above equation by re-ordering terms and using the fact that the average elastic response vanishes, and we obtain an equation for the pressure $p$ at short time scales 
\begin{equation}
\begin{split}
    p =   p_{\rm ext} + \frac{E}{1-\nu^2} \frac{\left( a - \langle a \rangle \right)}{e}  - \sigma_{\rm active} + \frac{1}{a} \left\langle   a \left(p - p_{\rm ext} + \sigma_{\rm active} \right) \right\rangle.
    \label{eq:forcebalance3}
\end{split}
\end{equation}
In addition, we will see below in Sec.~\ref{sec:secLongTimes} that the time-averaged term on the second line in Eq.~\eqref{eq:forcebalance31} is constant at short time scales. 

Eq.~\eqref{eq:forcebalance3} allows us to characterize the pressure in the system. In a single tube, for example, this would be a necessary equation to completely solve the fluid flow problem since we have a priori four unknowns, $v_r$, $v_z$, $p$, and $a$, and the Navier-Stokes equations only give three equations. Eq.~\eqref{eq:forcebalance3} relating pressure to circumferential stress and active stresses (or variants) has been used by several authors~\cite{shapiro1977steady,grotberg2004biofluid,mikelic2007fluid,elbaz2016axial,acosta2017cardiovascular,alim2017mechanism,julien2018oscillatory} (for example see Eq. (1c) of \cite{grotberg2004biofluid} or Eq.~(3) in \cite{shapiro1977steady}).
Eq.~\eqref{eq:forcebalance3} together with the hydrodynamic Eqs.~\eqref{eq:flows}-\eqref{eq:massConservation}, now form a complete set of equations to calculate flows and contractions at \textit{short} timescales. Our interest goes beyond, to the long timescales, where significant vein adaptation happens.

\subsection{Long timescales}
\label{sec:secLongTimes}

Let us now return to the full force balance Eq.~\eqref{eq:forcebalance2} and average dynamics over the short timescales to obtain long timescales. We will now approximate, for any variable $X$ and $Y$, $\langle X . Y \rangle \simeq \langle X \rangle \langle Y \rangle$ as we expect short timescale variations are small compared to long timescale variations $|X- \langle X \rangle|/\langle X \rangle \ll 1$. This is especially true in \pp~for radius dynamics where radius values oscillate periodically by about $5~\mathrm{\mu m}$ over short time scales of $1-2~\mathrm{min}$ while vascular adaptation ranges typically up to $50 \mathrm{\mu m}$ on long time scales of $10-30~\mathrm{min}$~\cite{marbach2021network}. Using the approximation $\langle X . Y \rangle \simeq \langle X \rangle \langle Y \rangle$ in the time-averaged Eq.~\eqref{eq:forcebalance2}, we obtain
\begin{equation}
    \frac{d\langle a\rangle}{dt} = \frac{2 \pi \langle a \rangle}{\gamma} \left(\langle (p - p_{\rm ext}) \rangle + 0  + \langle \sigma_{\rm active} \rangle + \mu \frac{ \langle \tau \rangle^2}{\tau_c} \right), 
    \label{eq:forcebalance4}
\end{equation}
where we used the fact that the average elastic response vanishes. 

\begin{figure}
\centering
\includegraphics[width=0.75\textwidth]{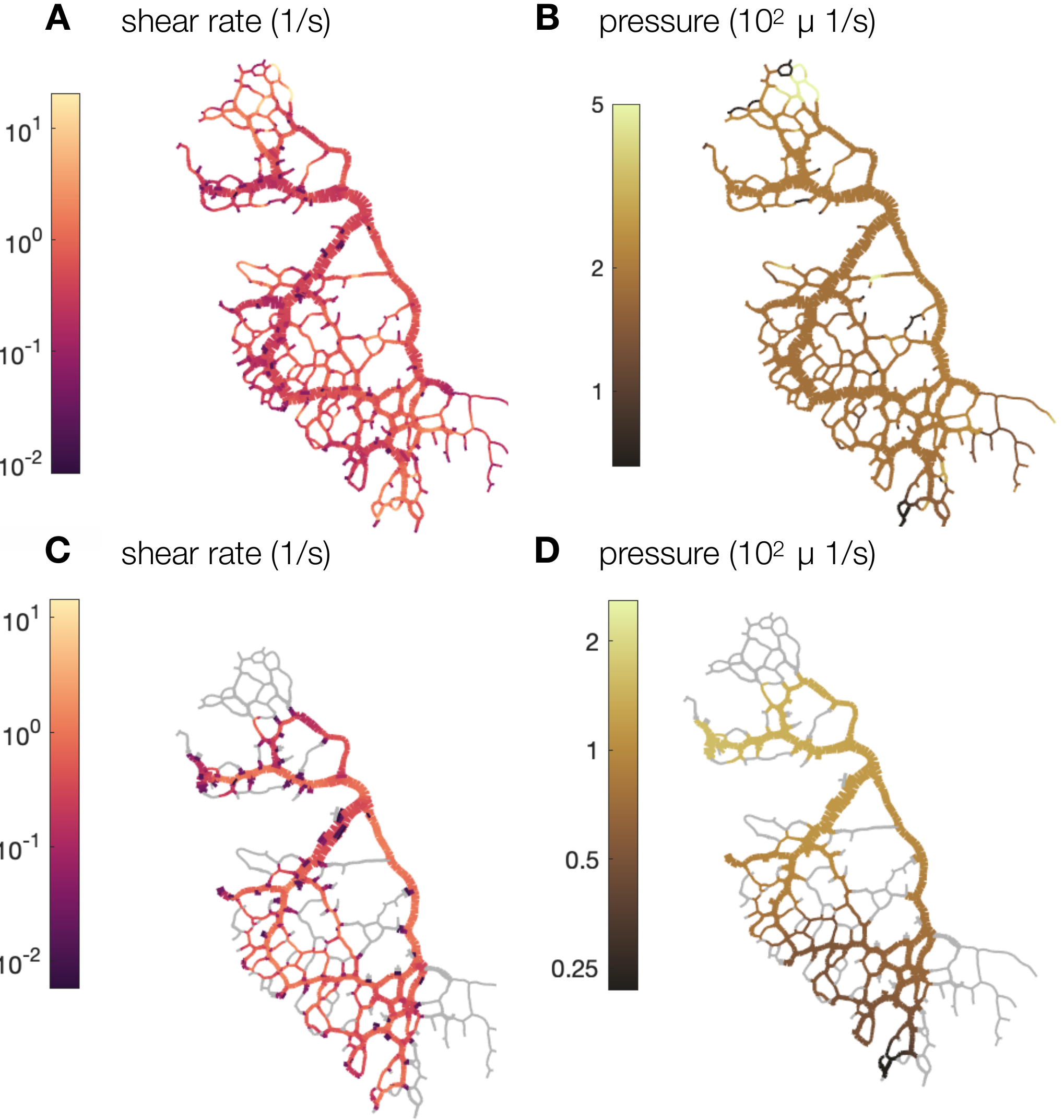}
\caption{Averaged shear rate (A-C) and pressure (B-D) in a specimen corresponding to the full network presented in Fig.~\ref{fig:intro}. The fields are calculated using data from image analysis that provides vein contractions, as well as conservation of mass and Kirchhoff's laws. The results are presented at the initial stage (A-B) and the final stage of the experiment, $54~\text{min}$ later (C-D). In the final stage, veins that have vanished are represented in gray. More details on calculation methods may be found in Ref.~\cite{marbach2021network}. Images are adapted with permission from Ref.~\cite{marbach2021network}.}
\label{fig:pressureShear}
\end{figure}
We now seek possible simplifications on the right-hand side of Eq.~\eqref{eq:forcebalance4}. 

Firstly, we can consider that $\langle (p - p_{\rm ext}) \rangle$  changes only marginally over long times. Generally, we expect that local pressure can not vary too much at the risk of damaging tissue. In \pp~we can verify this assumption by observing calculated pressure fields. In Fig.~\ref{fig:pressureShear}-B and D, we present numerically calculated pressure fields in each vein from network-wide vein contraction data for the specimen shown in Fig.~\ref{fig:intro}, at the beginning of the experiment and after a significant adaptation time. The details of how the pressure field is calculated are reported in Ref.~\cite{marbach2021network}. We find that the pressure field evolves only over typically one order of magnitude over the entire duration of the experiment of several hours. In comparison, shear can vary over three orders of magnitude in the same time frame; see Fig.~\ref{fig:pressureShear}-A and B. Our maps also show that the pressure $p$ is relatively uniform across the network and does not depend on the radius $\langle a \rangle$ of a vein, unlike shear rate. We can therefore consider that $\langle (p - p_{\rm ext}) \rangle$ smoothly evolves across the network and only changes over ``very'' long timescales,  longer than the timescale for a local vascular adaptation event. 

Secondly, as we mentioned earlier, $\langle \sigma_{\rm active} \rangle$ is a constant stress related to the active consumption of energy for wall contractility. Typically, as this stress characterizes the force per unit wall area necessary for contraction, we do not expect it to depend on $\langle a \rangle$. Importantly though, this active stress is related to energy consumption and hence constantly opposes vein growth; mathematically $\langle \sigma_{\rm active} \rangle < 0$, it can be thought of as the force analog of the metabolic cost energy $\pi b L a^2$ in Murray's derivation in Eq.~\eqref{eq:Murray_energy}, see the introduction. We, therefore, expect this component to determine the global sign of $\langle (p - p_{\rm ext}) + \sigma_{\rm active} \rangle$ such that we can write $\langle (p - p_{\rm ext}) + \sigma_{\rm active} \rangle = - \mu {\tau}_{\rm target}$ where $\tau_{\rm target} >0$ is a typical target shear rate, characterizing energy consumption, corrected by hydrostatic pressure, that can smoothly evolve across the network and over long timescales, longer than the timescale for local vascular adaptation. 
Under these conditions
Eq.~\eqref{eq:forcebalance4} simplifies to 
\begin{equation}
    \frac{d\langle a\rangle}{dt} = \frac{2 \pi \langle a \rangle}{\gamma} \left(\mu \frac{ \langle \tau \rangle ^2}{\tau_c} - \mu \tau_{\rm target} \right) ,
    \label{eq:forcebalance5}
\end{equation}
To further simplify this expression, we now introduce the local target shear rate $\tau_{0} = \sqrt{\tau_c {\tau}_{\rm target}}$ and
\begin{equation}
    t_{\rm adapt} = \frac{\gamma}{2\pi \mu \tau_{\rm target}}
\end{equation}
as characteristic adaptation timescale for vascular rearrangement. 

We, thus, rewrite Eq.~\eqref{eq:forcebalance5} in a more compact way and obtain the adaptation rule 
\begin{equation}
    \frac{d\m a}{dt} = \frac{\m a}{t_{\rm adapt}} \left(\frac{\langle \tau \rangle^2}{\tau_0^2} - 1  \right).
    \label{eq:forcebalance6}
\end{equation}
as the main result of this work. Note, Eq.~\eqref{eq:forcebalance6} resembles previous phenomenological approaches ~\cite{Taber1998Model,hacking1996shear,hu2012blood,tero2007mathematical,akita2016experimental, baumgarten2013functional,ronellenfitsch2016global,pries1998structural,pries2005remodeling,secomb2013angiogenesis,hu2013adaptation} yet, here, arises from force balance on the vein wall. 

\section{Discussion}

In this work, we have established a physical derivation based on force balance to justify the broadly used adaptation rule
\begin{equation}
    \frac{1}{\langle a \rangle} \frac{d \langle a \rangle}{dt} = \frac{1}{t_{\rm adapt}} \left( f(\tau) - f(\tau_0) \right).
    \label{eq:adaptationEND}
\end{equation}
For our case study in \pp, where fluid flow is encapsulated by crosslinked fibers making up a gel, we have provided reasoning to support that $f(\tau) = \langle \tau \rangle^2/\tau_0^2$ is in agreement with experimental data~\cite{janmey2007negative}. We recall that $a$ is the radius of a vein, $t$ time, $t_{\rm adapt}$ a timescale characterizing adaptation, $\tau$ shear rate, and $\tau_0$ a steady state shear rate. The notation $\langle . \rangle$ averages out short time scale, elastic deformations, such that the law in Eq.~\eqref{eq:adaptationEND} characterizes long timescale plastic deformations. 

Our force-balance approach is based on fluid flow physics and on the detailed enumeration and investigation of all the forces at play on the vessel wall, namely: hydrodynamic forces that are pressure and normal stress, potential forces such as circumferential stress, active stresses, anisotropic feedback from shear stress, and friction forces. We have shown that over long time scales corresponding to plastic deformations, the dominant forces are anisotropic feedback forces from shear stress. These are due to the unique feature of crosslinked fiber networks making up the actomyosin cortex. When the fibers are sheared, the crosslinks between them bring them closer together, resulting in a normal response under shear. This response tends to dilate vessels with larger shear rate. 

We now discuss our model's validity relative to other works and other systems beyond \pp, specific insight on Eq.~\eqref{eq:adaptationEND}, and possible extensions. 

\subsection{Validity of the derivation with respect to existing theories}

The result in Eq.~\eqref{eq:forcebalance6} has the same mathematical shape as the phenomenological Eq.~\eqref{eq:phenomenological} used in many prior works~\cite{Taber1998Model,hacking1996shear,hu2012blood,tero2007mathematical,akita2016experimental, baumgarten2013functional,ronellenfitsch2016global,pries1998structural,pries2005remodeling,secomb2013angiogenesis,hu2013adaptation}. It is therefore consistent with these phenomenological laws while bringing physical validation. Our result is also consistent with Murray's steady state assumption. In fact, the steady state of Eq.~\eqref{eq:forcebalance6} corresponds to a constant average shear rate in the vein $\langle \tau \rangle = \tau_0$.
Compared to existing theories where short timescale elastic deformation of veins is not discussed, here we provide a distinction between plastic and elastic deformations, and our adaptation rule corresponds to plastic deformations. The relevant or sensed shear rate for adaptation feedback is the long timescale one, $\langle \tau \rangle$, where short-lived elastic contributions are averaged out. 

\subsection{Functional dependence of the feedback on shear rate $\tau$}

In our derivation, we obtain that the adaptation function is quadratic, $f(\tau) =\frac{\langle \tau \rangle^2}{\tau_0^2}$. This is similar to the functional form used in Ref.~\cite{hu2013adaptation}, where the quadratic dependence was obtained phenomenologically from Murray's law. However, the origin of the quadratic dependence lies in the detailed characteristics of the anisotropic response of the gel, namely $f(\tau) = \frac{\sigma_r{\mu \langle \tau \rangle}}{\mu \tau_c}$. Hence, different mechanical properties of the gel making up a vein wall could yield different functional forms~\cite{conti2009cross}. Again, we argue that here it is not so much the exact functional dependence that is critical to obtain an adaptation model in the form of Eq.~\eqref{eq:phenomenological}. Rather, the fact that the gel making up the wall exhibits an anisotropic response is key in the force balance approach to obtain radius adaptation from  variable shear rate. 
%
%
%

\subsection{New insights from the force balance perspective}

The advantage of the force balance approach is that we can now give further physical meaning to the quantities $t_{\rm adapt}$ and $\tau_0$ in the adaptation law Eq.~\eqref{eq:phenomenological}.

The constant $\tau_{0} \sim - \langle \sigma_{\rm active} \rangle /\mu - \langle (p-p_{\rm ext}) \rangle/\mu $ corresponds to the steady state shear rate in Murray's law. It can, thus, be related to the typical local energy expense to sustain a vein $\sqrt{b/\mu}$, where we recall that $b$ is a local metabolic constant per unit volume and $\mu$ fluid viscosity. This contribution corresponds in our derivation to the active stress required to sustain peristaltic contractions $- \langle \sigma_{\rm active} \rangle /\mu \sim \sqrt{b/\mu}$. Here, we bring further insight complementing Murray's derivation, as our adaptation dynamics Eq.~\eqref{eq:forcebalance4} hints that $\tau_{\rm 0}$, or the metabolic cost, also depends on local pressure ($\langle p - p_{\rm ext}\rangle$). Hence, the local target shear rate $\tau_0$ is not just an intrinsic property of the system. Instead, it characterizes minimal energy expense at a given point in the network and smoothly and slowly varies across the network. Interestingly, our approach allows us to integrate the role of hydrostatic pressure in adaptation: when pressure is higher, the constant $\tau_0$ is decreased, favoring veins with \textit{ab initio} lower shear rate $\langle \tau \rangle$ to grow, in line with the physical intuition that high hydrostatic pressures may drive vein dilation. 

We can also draw insight on the adaptation timescale $t_{\rm adapt} = \frac{\gamma}{2\pi \mu \tau_{\rm 0}}$ as it includes the parameters of the model. If the local target shear rate $\tau_{0}$ is small, corresponding to a lower energy consumption level or larger local pressure that helps to keep the vein open, then $t_{\rm adapt}$ is long, and the vein is not prone to fast vascular adaptation. Reversely, if $\tau_{0}$ is large, vein adaptation can happen fast. Furthermore, the adaptation is slow coherently if the resistance to plastic change $\gamma$ is large. 

Finally, it is important to note that $t_{\rm adapt}$ and $\tau_0$ form two independent parameters characterizing the adaptation dynamics in Eq.~\eqref{eq:forcebalance6}, that both vary smoothly across the network. In fact, 
the parameters that define $t_{\rm adapt}$ and $\tau_0$, namely $\gamma$, $\mu$, and $\tau_c$ depend on mechanical and fluidic properties that vary across an individual organism as a function of both vein maturation and size \cite{swaminathan1997photobleaching,puchkov2013intracellular,Fessel.2017,Lewis.2015} as well as integrated exposure to light \cite{Bauerle2019}. The parameters $\gamma$, $\mu$ and $\tau_c$ also vary among different specimens due to the responsiveness to ambient conditions, such as humidity \cite{rakoczy1973myxomycete, takahashi1997asymmetry}, light conditions \cite{rakoczy1973myxomycete, hato1976phototaxis, nakagaki1996action, rodiek2015migratory} and temperature \cite{wohlfarth1977oscillating, hejnowicz1980propagated}. For example, the cytoplasm viscosity $\mu$ can vary depending on the local content of salt concentration or dispersed particles inside veins~\cite{puchkov2013intracellular}. Furthermore, both $\gamma$ and $\tau_c$ are related to the cortex mechanical properties, whose structure varies both within a specimen and over time \cite{Fessel.2017,Lewis.2015}. 

\subsection{Comments on extensions of the model and conclusion}

In our adaptation model, we consider a section of the vein that communicates only via flow with the rest of the network. A more detailed model could describe, for example, the dynamics of the retraction phase of a vein from its dangling end. As these are typically extremely short events, achieved within less than a contraction period, compared to vein dynamics such as shrinking or growing that can extend over several contraction periods, we chose to ignore them in our long timescale adaptation model. Note, that also other feedback mechanisms exist, such as fiber resistance~\cite{taber1998optimization}, wall thickness adaptation~\cite{pries2005remodeling}, and more detailed processes in time such as energy or oxygen transport~\cite{secomb2013angiogenesis}. 

Here, the model derived from first principles incorporates as few assumptions as possible to arrive at the adaptation rule Eq.~\eqref{eq:phenomenological}. Using the quadratic dependence of the adaptation function $f(\tau) = \langle \tau \rangle^2/\tau_0^2$, we have shown in our accompanying work~\cite{marbach2021network}, that even these simple assumptions are sufficient to reproduce a variety of adaptation dynamics that are observed experimentally. Furthermore, although our force balance derivation and subsequent experimental investigation were adapted to the model organism \pp, the underlying physical principles of fluid flow physics and mechanical response are universal. 
Hence, we believe adaptation models based on force balance approaches are relevant to study vascular adaptation across further flow networks in plants and animals.

\section*{Acknowledgements}
The authors are indebted to Charles Puelz for enlightening discussions on force balance in veins and Emilie Verneuil for discussions on the anisotropic response of sheared gels. They would also like to thank Leonie Bastin and Felix B\"auerle for interesting discussions on \pp. S.M.~was supported in part by the MRSEC Program of the National Science Foundation under Award Number DMR-1420073. This work was supported by the Max Planck Society and has received funding from the European Research Council (ERC) under the European Union’s Horizon 2020 research and innovation program (grant agreement No. 947630, FlowMem).

%

%
%
%


\section*{References}

\providecommand{\newblock}{}

\end{document}